\documentclass[aps,prb,twocolumn,floatfix,showpacs,citeautoscript,superscriptaddress,longbibliography,hyperlinks]{revtex4-2}
\usepackage{amsmath}
\usepackage{amssymb}
\usepackage{amsfonts}
\usepackage[colorlinks=true,citecolor=blue]{hyperref}
\usepackage{graphicx}
\usepackage{enumitem}
\setlist[enumerate]{leftmargin=6mm}
\usepackage[caption = false]{subfig}
\usepackage{braket}
\usepackage{siunitx}
\usepackage{tikz}
\usepackage{bbm}
\usepackage{circuitikz}
\usepackage[capitalise]{cleveref}
\usepackage{booktabs}

\usepackage{glossaries}
\glsdisablehyper    
\newacronym{ti}{3dTI}{three-dimensional topological insulator}
\newacronym{nfn}{NFN}{normal-ferromagnet-normal}

\newcommand{\sii}{\hat{\sigma}_{0}}

\newcommand{\mbf}[1]{\mathbf{ #1 }}

\newcommand{\md}{\mathrm{d}}
\newcommand{\mi}{\mathrm{i}}
\newcommand{\me}{\mathrm{e}}

\definecolor{crimson}{RGB}{255,102,255}
\definecolor{mahogany}{RGB}{192,64,0}

\usepackage{sidecap,tikz}
\definecolor{lime}{HTML}{A6CE39}
\DeclareRobustCommand{\orcidicon}{\hspace{-1mm}
	\begin{tikzpicture}
		\draw[lime, fill=lime] (0,0) 
		circle [radius=0.16] 
		node[white] {{\fontfamily{qag}\selectfont \tiny \,ID}};
		\draw[white, fill=white] (-0.0525,0.095) 
		circle [radius=0.007];
	\end{tikzpicture}
	\hspace{-3mm}
}
\foreach \x in {A, ..., Z}{\expandafter\xdef\csname orcid\x\endcsname{\noexpand\href{https://orcid.org/\csname orcidauthor\x\endcsname}
		{\noexpand\orcidicon}}
}

\begin{document}

\newcommand{\uam}{Department of Theoretical Condensed Matter Physics, Universidad Aut\'onoma de Madrid, 28049 Madrid, Spain}
\newcommand{\ifimac}{Condensed Matter Physics Center (IFIMAC), Universidad Aut\'onoma de Madrid, 28049 Madrid, Spain}
\newcommand{\inc}{Instituto Nicol\'as Cabrera, Universidad Aut\'onoma de Madrid, 28049 Madrid, Spain}

\newcommand{\tianjin}{Department of Physics, Tianjin University, Tianjin 300072, China}

\newcommand{\nagoya}{Department of Applied Physics, Nagoya University, Nagoya 464-8603, Japan}

\newcommand{\duisburg}{Theoretische Physik, Universit\"at Duisburg-Essen and CENIDE, D-47048 Duisburg, Germany}

\title{Gate-tunable nonreciprocal thermoelectric effects on the surface states of topological insulators} 

\author{Phillip Mercebach\orcidF{}}
\affiliation{\uam}
\affiliation{\ifimac}

\author{Sun-Yong Hwang}
\affiliation{\duisburg}

\author{Bo Lu\orcidD{}}
\affiliation{\tianjin}

\author{Bj\"orn Sothmann\orcidC{}}
\affiliation{\duisburg}

\author{Yukio Tanaka\orcidB{}}
\affiliation{\nagoya}

\author{Pablo Burset\orcidA{}}
\affiliation{\uam}
\affiliation{\ifimac}
\affiliation{\inc}

\date{\today}

\begin{abstract}
Thermoelectric devices at the nanoscale offer promising routes for on-chip refrigeration and waste-heat recovery, yet most semiconductor-based implementations suffer from limited tunability and narrow operational ranges. We introduce a highly flexible thermoelectric platform based on a ballistic junction formed by two gate-tunable regions of a topological insulator surface state bridged by a magnetic barrier. We theoretically demonstrate that such device exhibits strong electrical control over both refrigeration and thermoelectric power generation via side gates. We exploit the interplay between strong spin-orbit coupling and magnetism to achieve pronounced nonreciprocal transport, asymmetric cooling and tunable diode-like behavior. 
To demonstrate experimental feasibility, we further analyze refrigeration efficiency and phonon-limited performance in realistic material settings. 
\end{abstract}

\maketitle
	
\section{Introduction \label{sec:intro}}
Controlling heat and charge transport at the nanoscale is a key challenge in modern condensed matter physics and nanotechnology~\cite{Li2012Jul}. Mesoscopic thermoelectric devices enable on-chip conversion of thermal gradients into electrical energy or localized electronic cooling~\cite{Roberts2011May}. 
Consequently, several nanoscale platforms have been investigated for targeted cooling and energy harvesting, which are essential for reducing thermal noise, managing dissipation, and improving device efficiency. 
Examples of such platforms are semiconductor quantum dots~\cite{Josefsson2018Oct,Nugraha2023Sep,Balduque2025May}, superconducting hybrid structures~\cite{Giazotto2006Mar,Germanese2022Oct,Karimi2018,Sanchez2018,Kirsanov2019Mar,Pekola2021Oct,Strambini2022May,Arrachea2025May}, molecular junctions~\cite{Xu2023Nov,He2025Mar,Preesam2024Nov}, bacteria~\cite{Krishnamurthy2016Dec}, 
and two-dimensional materials, such as graphene~\cite{Chang2014Jul,Dragoman2007,Alomar2014,*Alomar2014Err,Yokomizo2013}. Strong thermoelectric responses are achieved leveraging quantum confinement~\cite{Sothmann2014Dec}, Coulomb interactions~\cite{Yang2019Jul}, or broken symmetries~\cite{Mani2019,Mani20192,WONG2021120607}. 
Nevertheless, challenges related to low tunability and constrained operating conditions persist, especially in conventional semiconductor-based designs. 

Among the various platforms for mesoscopic thermoelectric devices, Dirac materials—and in particular, topological insulators~\cite{Ma2013Mar,Xu2017Sep,Gusev2021Dec,Alisultanov2025Apr,Fu2020Apr}—stand out due to their unique electronic properties~\cite{Wehling2014Jan}. 
Low-energy excitations in these materials obey a relativistic Dirac equation, leading to linear dispersion, suppressed backscattering, and high carrier mobility even at low temperatures~\cite{Wehling2014Jan}. 
In topological insulators, robust edge states with spin-momentum locking are protected by time-reversal symmetry, allowing for dissipationless charge transport along the edges or surfaces~\cite{Breunig2022}. 
These properties make topological insulators ideal candidates for energy-efficient thermoelectric applications, where strong coupling between heat and charge flow is essential and tunability via gating, doping or proximity effects can be readily implemented~\cite{Zhang2011,Uesugi2019,Grutter2021Sep,Wang2023Aug}. 
Furthermore, their linear dispersion and low density of states near the Dirac point enhance the sensitivity of transport to external fields, making them ideal for designing tunable thermoelectric devices~\cite{Mahfouzi2010Nov,Wang2014Mar,Li2023Feb,Shi2025Mar,Mukhopadhyay2025Jan,Zhang2025Jun}. 
These characteristics, combined with compatibility with two-dimensional material platforms and scalable device architectures, position \glspl{ti} as a versatile and technologically relevant choice for nanoscale heat and energy management.

\begin{figure}[b]
	\includegraphics[width=1.0\linewidth]{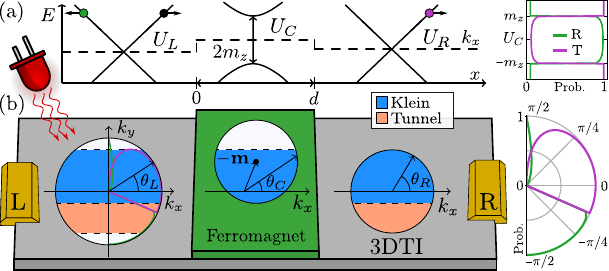}
	\caption{\gls{nfn} junction on the surface state of a \gls{ti}. 
		(a) Energy bands of each region at $k_y=0$ for out-of-plane magnetization $m_z$. Local potentials $U_{L,C,R}$ can be tuned by electric side gates. Right: transmission ($T$, magenta) and reflection ($R$, green) probabilities for an injected electron (black dot). 
		(b) Sketch of the setup indicating normal (gray) and magnetic (green, in-plane magnetization) regions, and their Fermi surfaces. The angular distribution of $T$ (magenta) and $R$ (green) is shown on the right and on the leftmost Fermi surface. 
	}\label{fig:setup} 
\end{figure}

In this work, we propose a thermoelectric device architecture based on the ballistic surface states of a \gls{ti}. By introducing a magnetic barrier between two gate-tunable normal regions, we create a \gls{nfn} junction that enables highly controllable thermoelectric and refrigeration performance (\cref{fig:setup}). 
The interplay between spin-momentum-locked surface states and the magnetic barrier generates pronounced nonreciprocal effects, including diode-like charge and heat transport and asymmetric cooling. We further assess the refrigeration efficiency and the effective electron cooling temperature, accounting for phonon contributions relevant to realistic quasi-two-dimensional materials such as graphene and topological insulators. 

The rest of the paper is organized as follows. 
We present the theoretical model and describe the main transport mechanisms responsible for the thermoelectric effects in \cref{sec:model}. 
Next, we explore the different thermoelectric effects in \cref{sec:current}. 
\Cref{sec:device} includes our analysis of the efficiency of the proposed platform for thermoelectric energy production and local cooling. 
We finish with our conclusions in \cref{sec:conc}. 
Further details on our calculations are given in the Appendix. 

\section{Model\label{sec:model}}

We consider a surface state of a \gls{ti} extending across the $x$-$y$ plane, see \cref{fig:setup}. 
A central magnetic region $C$ located at $0<x<d$ separates two metallic regions connected to left ($L$) and right ($R$) reservoirs with well-defined chemical potentials, $\mu_{L,R}$, and temperatures, $T_{L,R}$. We assume the system width along the $y$-direction to be much larger than the distance between reservoirs, and assume clean and flat interfaces that conserve the parallel momentum $\hbar k_y$. 
The low-energy electronic excitations with momentum $\hbar\mbf{k}=\hbar(k_x,k_y,0)$ are then described by the Dirac Hamiltonian~\cite{Linder_2009, Yokoyama2010Mar, Allain2011, Zhang2012Nov, Pau_2015}
\begin{equation}\label{eq: H}
	\hat{H} (\mbf{k})= \hbar v_F  \mbf{k} \cdot \mbf{\hat{\sigma}} - U(x) \sii + \mbf{m}(x) \cdot \boldsymbol{\hat{\sigma}}, 
\end{equation}
where $v_F$ is the Fermi velocity, $\sii$ is the identity matrix and $\boldsymbol{\hat{\sigma}}$ the vector of Pauli matrices $\hat{\sigma}_{x,y,z}$ acting on spin space, and
the electrostatic potential along the edge state is defined piecewise as $U(x)= U_L \Theta(-x) + U_C \Theta(x,d-x)+ U_R \Theta(d+x)$, with $\Theta(x)$ being the Heaviside function and $\Theta(x,y)=\Theta(x)\Theta(y)$. 
In experiments, local potentials $U_{L,C,R}$ can be independently tuned by electric side gates. 
The magnetization $\mbf{m}(x) = (m_x, m_y, m_z)^{T}\Theta(x,d-x)$ is only finite in the intermediate region $C$. 
The dispersion associated to \cref{eq: H} involves (massive) Dirac bands
\begin{equation}\label{eq:energy}
	E = - U_{j} \pm \left\| \hbar v_F \mbf{k} + \mbf{m}(x) \right\| ,
\end{equation} 
with $j=L,C,R$. 
Translation invariance along the $y$ direction allows us to write the eigenstates of \cref{eq: H} as $\Psi_{j\pm}(x, y)= \me^{\mi k_y y}\psi_{j\pm}(x,\theta_j)$, with 
\begin{equation}\label{eq:snell}
k_y=  \frac{ \left| E+U_{n}\right| }{\hbar v_F} \sin\theta_{n} = \frac{ \sqrt{ (E+U_C)^2 - m_z^2} }{\hbar v_F} \sin\theta_C.
\end{equation}
Here, $n=L,R$ labels only the external regions
where the solutions are Dirac spinors 
\begin{equation}\label{eq:spinor-N}
\psi_{n\pm}(x,\theta_{n})= \frac{1} { \sqrt{2\cos\theta_{n}} }\begin{pmatrix} 
		1 \\ \pm\me^{\pm \mi \theta_{n}}
		\end{pmatrix} \me^{\pm \mi k_{n} x}, 
\end{equation}
with 
\begin{equation}\label{eq:wavenumber-N}
		k_{n} = \mathrm{sgn} (E + U_{n}) \sqrt{(E+U_{n})^2/(\hbar v_F)^2 - k_y^2 }  ,
\end{equation}
and the angles $\theta_{n}$ defined according to \cref{eq:snell}. 
On the central region the solutions take the form
\begin{equation} \label{eq:spinor-C}
	\psi_{C\pm}(x, \theta_C) = \frac{ \lambda_{2} \me^{-\mi m_x x/(\hbar v_F)} }{ \mathcal{N} \lambda_{1} } 
	\begin{pmatrix} 
		1 \\ \pm 
		\me^{\pm \mi \theta_C} 
		\end{pmatrix} 
		\me^{\pm \mi k_C x}, 
\end{equation}
with $\lambda_{1,2}= \sqrt{|E+U_C \pm m_z|}$, $\mathcal{N}$ being the normalization constant and 
\begin{equation}\label{eq:wavenumber-C}
	k_{C} = \frac{\mathrm{sgn} (E + U_{C})} {\hbar v_F} \sqrt{ \lambda_{1}^2\lambda_{2}^2 - (\hbar v_F k_y + m_y)^2} ,
\end{equation}
where we have defined 
\begin{equation}\label{eq:extra-defs}
\me^{\pm \mi\theta_C} = \gamma \frac{ \hbar v_F k_C \pm \mi \left( \hbar v_F k_y + m_y \right) }{ \lambda_{1}\lambda_{2} }, 
\end{equation}
with $\gamma= \mathrm{sgn}(E + U_C + m_z)$. 

The scattering problem is defined matching scattering states of the type $\phi_j(x,\theta_j)= \sum_{\alpha=\pm} a_{j\alpha} \psi_{j\alpha}(x,\theta_j)$ between the magnetic and metallic regions, i.e.,  
$\phi_L(0,\theta_L)=\phi_C(0,\theta_C)$ and $\phi_C(d,\theta_C)=\phi_R(d,\theta_R)$~\cite{Linder_2009, Yokoyama2010Mar, Allain2011, Zhang2012Nov, Pau_2015}. By considering injection from one of the metallic regions, e.g., $a_{L+}=1$ and $a_{L-}=0$ for the left lead, one obtains the transmission amplitudes $t_{RL}(E,\theta)=a_{R+}$ (in this case $a_{R-}=0$). 
Using \cref{eq:snell}, all angles can be written in terms of the angle of incidence $\theta$ defined in one of the external regions, i.e., $\theta=\theta_L$ ($\theta=\theta_R$) when incoming particles originate from lead $L$ ($R$). 
The resulting tunneling conductance is the normalized sum over all channels of the transmission probability $T_{RL}(E,\theta)= \mathrm{Re}\{ k_R/k_L \} |t_{RL}(E,\theta)|^2$ [analogously for $T_{LR}(E,\theta)$, see \cref{sec:app-A} for more details]~\cite{Linder_2009, Yokoyama2010Mar, Allain2011, Zhang2012Nov, Pau_2015},
\begin{equation}\label{eq:conductance}
	\sigma_{n}(E) = \sigma_0 \left| E+U_{n} \right| \int_{-\pi/2}^{\pi/2} T_{\bar{n}n}(E,\theta) \cos(\theta) \md \theta, 
\end{equation}
where $\bar{n}=R,L$ when $n=L,R$, and $\sigma_0 = W / (2\hbar v_F)$, with $W$ being the junction width. 
The charge and heat currents are then given by~\cite{Kheradsoud_2019,Pau_2020,Rafa_2020}
\begin{subequations}\label{eq:current}
	\begin{align}
		I_{n} ={}& \frac{2e}{h} \int_{-\infty}^{\infty} \md E \sigma_{n}(E) \delta f_{n\bar{n}}(E) , \\
		J_{n} ={}& \frac{2}{h} \int_{-\infty}^{\infty} \md E \left( E- eV_{n}\right)\sigma_{n}(E) \delta f_{n\bar{n}}(E) ,
	\end{align} 
\end{subequations}
where $\delta f_{n\bar{n}}(E)= f_{n}(E) - f_{\bar{n}}(E) $, with $f_n(E) = 1/\left\{1+\exp\left[(E-eV_n)/(k_B T_n)\right]\right\}$ being the Fermi function for reservoir $n=L,R$ at temperature $T_{n}$ ($k_B$ is the Boltzmann constant) and applied voltage $eV_n=\mu_n-\mu_0$, measured with respect to the equilibrium chemical potential $\mu_0=0$. In the following, we normalize the currents by $J_0=2\sigma_0/h$ and $I_0=eJ_0$. 

\begin{figure}[t]
	\includegraphics[width=1.0\linewidth]{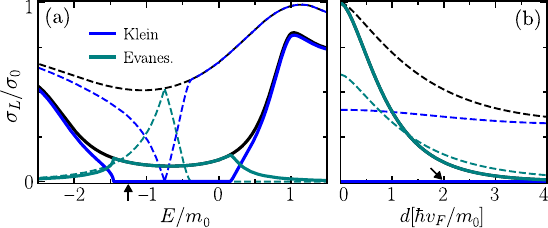}
	\caption{Conductance on the left lead (black lines), and the contribution from Klein (blue lines) and evanescent (green lines) processes, for $m=0$ (dashed lines) and $m = m_0$ (solid lines). 
	(a) Conductance as a function of the energy for $d=2 \hbar v_F/m_0$ and (b) as a function of $d$ for $E=-U_C-m_0/2$, marked by the arrow in (a). The rest of parameters are $\beta=\pi/4$ and $U_C/m_0=0.75$. 
	}
	\label{fig:scattering}
\end{figure}

The asymmetry in transport processes that we analyze below originates from the interplay between the intermediate magnetic region and the strong spin-orbit coupling of the surface state~\cite{Garate2010Apr,Burkov2010Aug,Schwab2011Mar,Yokoyama2011Apr,Nagaosa2012,Semenov2012Oct,Yokoyama2014Jan,Taguchi2014Feb}. 
The magnetization components parallel to the spin-orbit coupling of the surface state, $m_{x,y}$ in our notation, shift the momentum as described in \cref{eq:energy}~\cite{Linder_2009, Yokoyama2010Mar, Allain2011, Zhang2012Nov, Pau_2015}. Consequently, the magnetization in the direction of transport $m_x$ becomes an irrelevant phase factor in \cref{eq:spinor-C} and we can ignore it henceforth. 
By contrast, the magnetization along the $y-z$ plane, $\mbf{m} = m (0,\sin\beta,\cos\beta)$ with $m^2=m_y^2+m_z^2$, opens a gap in the junction conductance through two different, but complementary, mechanisms. 
Henceforth, we normalize all energy scales to a finite magnetization $m_0/(\hbar v_F k_F)=1$, even when we consider $m=0$. 

First, the magnetization orthogonal to the spin-orbit coupling, $m_z$, becomes a mass term in \cref{eq: H} that opens a gap in the spectrum of the central region, see \cref{fig:setup}(a). The position of the gap is controlled by $U_C$, inducing thermoelectric effects when $U_C\neq0$. 

By contrast, the magnetic component along the $y$ direction does not open a gap in the bulk bands, just shifts the position of the Dirac point in momentum space, see \cref{fig:setup}(b). Only when we consider transport across this shifted region we can observe a gap in the conductance (\cref{fig:scattering}). 
Additionally, the position of this $m_y$-conductance gap (for $m_z=0$) depends on the electrostatic potentials $U_j$ and can thus be manipulated electrically by side gates (\cref{fig:scattering}). 
Indeed, when considering electron transport from $L$ to $R$ as sketched in \cref{fig:setup}(b), the size of the Fermi surface at a given energy is controlled by each region's gating. Then, due to the Dirac point shift by $m_y$ only a wedge of the incident modes (blue area) finds available states in the central region. Electronic states with angles within that slice can thus transfer to $R$ via Klein scattering~\cite{Allain2011}. 
The rest of incident modes that can be matched to available states on $R$ (gray area) must tunnel through a gapped central region where the solutions of \cref{eq: H} become evanescent waves. Of course, incident modes that can not be connected to states in $R$ must backscatter (white area) due to Fermi vector mismatch. 

\Cref{fig:scattering}(a) shows that the conductance gap due to Klein processes (blue line) features hard edges, which help achieve strong thermoelectric effects as we show below. 
Evanescent processes (green line), by contrast, soften the total conductance gap (black line) and are thus a detrimental effect~\cite{Whitney_2015}. Fortunately, Klein scattering is not affected by the barrier width $d$, while tunneling through evanescent states decays exponentially with it, see \cref{fig:scattering}(b) for the width dependence of the conductance at a fixed energy. Consequently, we can suppress the effects of evanescent processes by adjusting the width of the central region. 

When both $m_z$ and $m_y$ are finite, the effects above (spectrum gap from $m_z$ and transport gap by $m_y$) combine in a nontrivial manner, see \cref{sec:app-B}. 
For example, the transport gap associated to $m_y\neq0$ is widened by setting $U_g\equiv U_L-U_R \neq 0$. At the same time, scattering across the junction becomes very asymmetric when $U_g-U_C \neq 0$. Therefore, as we show next, the effective conductance gap can be electrically controlled through the gates $U_g$ and $U_C$ for a generic orientation of the magnetization. 

\begin{figure*}[t]
	\includegraphics[width=1.0\linewidth]{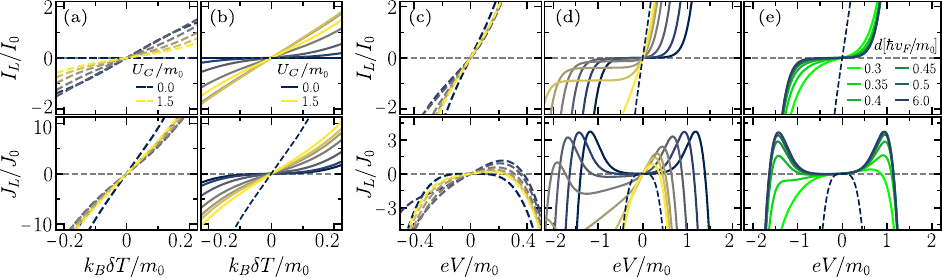}
	\caption{Charge (top panels) and heat (bottom panels) currents on the left lead for $m=0$ (dashed lines) and $m = m_0$ (solid lines). 
		(a,b,c,d) Currents as a function of (a,b) a temperature gradient $\delta T$ at zero bias, or (c,d) a voltage bias $eV$ at thermal equilibrium, for $d=6 \hbar v_F /m_0$ and different values of the central gate, $U_C/m_0=0$ (blue), 0.25, 0.50, 0.75, 1.00, 1.25, and 1.50 (yellow). 
		(e) Voltage dependence of the currents for different junction lengths $d$ at fixed $U_C/m_0=0.5$. Black dashed lines correspond to $m=U_C=0$ (gray dashed lines mark zero current). In all cases, $k_BT/m_0=1/8$, $U_L=U_R=0$, $T_{L,R}= T \pm \delta T$. 
	}
	\label{fig:currents}
\end{figure*}

\section{Thermoelectric effects\label{sec:current}}

We now analyze how the asymmetric transport through the magnetic barrier impacts the charge and heat currents across the junction, leading to strong non-linear effects in both the charge and heat currents. These nonlinearities lead to a thermoelectric diode effect that can be tuned electrically by gating the central region. 

\subsection{Thermoelectric current}

First, we consider a temperature gradient in the absence of applied voltage, i.e., when $T_L\neq T_R$ and $\mu_L=\mu_R$. 
For simplicity, we set $T_L=T+\delta T$ and $T_R=T-\delta T$, and only consider the transport asymmetry coming from electrically gating the central region while keeping $U_L=U_R=0$, \cref{fig:currents}(a,b). 
For $m=U_C=0$, there is no transport asymmetry and the electric current is zero, while the heat simply flows from hot to cold reservoirs [blue dashed lines in \cref{fig:currents}(a)]. Even in the absence of magnetic barrier, $m=0$, gating the central region ($U_C\neq 0$) leads to a sizable transport asymmetry resulting in a finite and mostly linear thermoelectric current [\cref{fig:currents}(a)]. 
Such a nonmagnetic thermoelectric effect originates from the asymmetry in transport induced by the different charge neutrality points between the central and the $L$ and $R$ regions (dashed lines in \cref{fig:scattering}). 
As such, the resulting electric current is mostly determined by evanescent processes and is, therefore, much more dependent on the length of the central barrier than the current for $m\neq 0$ [\cref{fig:currents}(b)]. 
This is a direct consequence of the magnetic region opening a gap in the conductance (\cref{fig:scattering}). 

In general, we observe a thermoelectric current, even for $m=0$, if the gating at the normal regions is different, $U_L\neq U_R$ ($U_g\neq 0$), or if we set $U_C\neq 0$ when $U_L=U_R$. 
In the presence of a magnetic barrier ($m \neq 0$), and for $U_C\simeq 0$, the currents become strongly nonlinear by being suppressed for small gradients comparable to the magnetic gap [blue solid lines in \cref{fig:currents}(b)]. 
Additionally, the current can be further increased by setting $U_C\neq0$ since, as explained above, the local gate $U_C$ at the magnetic barrier induces a strong asymmetry in transport. 
As $U_C$ approaches $m$, the linear behavior of both charge and heat currents is recovered. 
Consequently, the \gls{nfn} junction on the surface of a \gls{ti} acts like a strong heat engine that can be controlled electrically by side gates. We explore the optimal operation of the junction in \cref{sec:device}. 

\subsection{Nonreciprocal charge and heat currents}

Next, we consider a voltage-biased junction at thermal equilibrium ($T_L=T_R$) so, for simplicity, we fix $\mu_R=0$ and let $\mu_L\equiv eV\neq 0$. 
The charge current in the absence of magnetic barrier ($m=0$), black dashed line in \cref{fig:currents}(c), displays a typical Ohmic behavior where the current depends linearly on the applied voltage $V$, with slope given by the conductance. 
For a finite magnetization [\cref{fig:currents}(d)], the electric current is suppressed for voltages smaller or similar to the magnetic gap size, roughly given by $m$. 
Such suppression is symmetric around zero voltage for $U_C=0$ (blue line), but it becomes strongly asymmetric when $U_C\neq 0$ (purple and red lines). As $U_C$ becomes similar to $m$, the asymmetry in the charge current is increased, resulting in a diode effect for voltages comparable to $m$. 
By setting $d/(m_0/\hbar v_F)=6$ in \cref{fig:currents}(c) we have suppressed the detrimental effects from evanescent processes. For shorter junctions, the electric current is still very nonreciprocal, but it is no longer suppressed for negative voltages [\cref{fig:currents}(e)]. 

The magnetic barrier is also responsible for a cooling effect at voltages around the magnetic gap $eV\sim m$, see \cref{fig:currents}(d). 
Cooling manifests as a sign reversal in the heat current indicating that heat flows from the cold to the hot electrode ($J_L>0$) instead of the usual hot-to-cold flow ($J_L<0$). 
We first note that a small Peltier cooling is possible even for $m=0$, see black dashed lines in \cref{fig:currents}(d). 
As before, this effect originates from the asymmetry induced when the charge neutrality points from the Dirac spectrum at each region are different. Again, the currents generated are much smaller than the ones for $m\neq0$ because the semimetallic Dirac spectrum with $m=0$ does not feature a gap. 

Interestingly, by setting $U_C\neq 0$ and tuning it towards $|U_C|\sim m$, we also observe a strong asymmetry in the cooling peaks, reaching a situation where a positive bias immediately induces cooling while there is no heat flow for a wide range of negative voltages (light purple and red lines). 
Here, we have taken $U_g=0$ ($U_L=U_R=0$) so that a finite $U_C$ is enough to have $U_g-U_C\neq0$ and thus introduce a strong asymmetry on transport. For simplicity, we are also considering that transport is completely dominated by Klein processes, i.e., $d=2\hbar v_F/m$. We show in \cref{fig:currents}(g) the detrimental effect of evanescent processes in the asymmetric cooling. 

\begin{figure}[b]
	\includegraphics[width=1.0\linewidth]{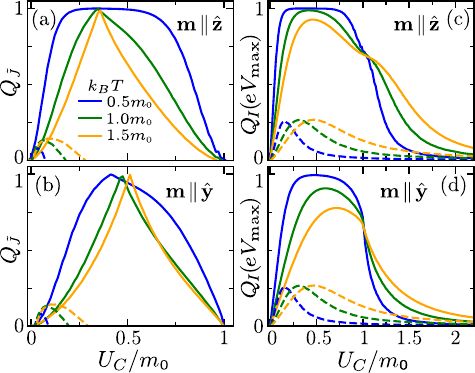}
	\caption{Diode quality factors for different configurations. 
		(a,b) Heat quality factor $Q_{\tilde{J}}$ and (c,d) charge quality factor $Q_I(eV)$, evaluated at $eV=eV_\text{max}$, as a function of $U_C$ for (a,c) $\mbf{m} \parallel \hat{ \mbf{z} }$ ($\beta=0$) and (b,d) $\mbf{m} \parallel \hat{ \mbf{y} }$ ($\beta=\pi/2$). 
		Solid (dashed) lines correspond to $m=m_0$ ($m=0$). 
	}
	\label{fig:diode}
\end{figure}

\subsection{Thermoelectric diode effect\label{sec:diode}}

We have established that the highly asymmetric transport of the \gls{ti} edge states across a magnetic barrier results in strong nonreciprocal effects in the charge and heat currents. We now quantify such a thermoelectric diode effect defining a quality factor $Q$. 

For the heat current we are interested in the maximum cooling power at thermal equilibrium $\tilde{J}_{\pm}\equiv J_L(\pm eV_\text{max}, \delta T=0)$, with $eV_\text{max}>0$ the voltage that maximizes the heat current, cf. \cref{fig:currents}(d). 
We can thus define a heat current diode quality factor as~\cite{Roberts2011May}
\begin{equation}\label{eq:diode-JV}
	Q_{\tilde{J}} = \frac{ |\tilde{J}_{+}| - |\tilde{J}_{-}| } { |\tilde{J}_{+}| + |\tilde{J}_{-} | } .
\end{equation}
Note that in a two-terminal setup we can focus on one of the leads to analyze the cooling power. 

We plot $Q_{\tilde{J}}$ in \cref{fig:diode} for $m\neq0$ (solid lines) at different temperatures, and compare it to the nonmagnetic case with $m=0$ (dashed lines). 
We consider an out of plane magnetization ($\mbf{m}=m \hat{\mbf{z}}$) in \cref{fig:diode}(a) and an in-plane one ($\mbf{m}=m \hat{\mbf{y}}$) in \cref{fig:diode}(b), although the results are qualitatively the same for both orientations. 
The cooling non-reciprocity in the nonmagnetic case is small ($Q_{\tilde{J}}(m=0)\lesssim10\%$) and quickly disappears for $U_C\neq0$. Having set $eV_\text{max}>0$, the sign change in $Q_{\tilde{J}}$ indicates that the Joule heating current ($J_L<0$) becomes dominant over the small cooling effects, cf. \cref{fig:currents}(c). 

By contrast, the magnetic case at low temperatures compared to the magnetic gap ($k_BT/m\ll1$) presents a finite $Q_{\tilde{J}}$ with values close to $1$ as long as $0<|U_C|<m$; the range being larger for the out of plane magnetization that features a harder gap.  Consequently, one reservoir is cooled for positive bias while there is roughly no heat transfer for the opposite voltage. 
The asymmetric cooling effect remains perfect ($Q_{\tilde{J}} =1$) at higher temperatures when the magnetic barrier is gated to a value around half of the effective magnetization, until it eventually vanishes for $k_BT\gg m$. The diode effect on the Peltier cooling disappears for $|U_C|>m$. 

Analogously, we can measure the asymmetry in the charge transport from a voltage bias. 
To do so, we define the charge diode quality factor 
\begin{equation}\label{eq:diode-IV}
	Q_{I} (eV) = \frac{ |I_L(eV)| - |I_{L}^{}(-eV)| } {|I_{L}^{}(eV)| + |I_{L}^{}(-eV)|} ,
\end{equation}
always assuming $T_L=T_R$. 
We plot $Q_I(eV_\text{max})$ in \cref{fig:diode}(c) [\cref{fig:diode}(d)] for out of plane (in-plane) magnetization using solid lines. 
As before, the dashed lines correspond to the $m=0$ case for comparison. 
We choose the same voltage $eV_\text{max}$ that maximizes the heat current to fully capture the non-reciprocity of the electric current. Indeed, the asymmetric charge transport could be visualized evaluating $Q_{I}$ at a voltage bigger than the magnetic gap size, since below it the current is suppressed, see \cref{fig:currents}(d). 
As expected from the analysis of the charge current above [top panels of \cref{fig:currents}(c) and (d)], for any voltage we have $Q_{I}(eV)=0$ for $U_C=0$ since, in this situation, the current is symmetric with the voltage. As we set $U_C/m_0\rightarrow 1$ the asymmetry in charge transport emerges and $Q_{I}(eV_\text{max})$ quickly becomes finite and close to $1$, specially for the out-of-plane magnetization [\cref{fig:diode}(c)]. 
Here, the \textit{hardness} of the gap, i.e., how sharp is the transition from zero to finite conductance, is very important. Consequently, evanescent processes for short junctions are thus very detrimental, and the spectrum gap from $m_z$ displays better quality factors than the transport gap from $m_y$.

\section{Thermoelectric device\label{sec:device}}

The asymmetric scattering of the \gls{ti} surface states through the magnetic barrier results in strong thermoelectric effects and nonreciprocal heat and charge currents out of equilibrium. 
We now consider how efficiently these currents can cool down a hot reservoir or generate electric power out of a temperature gradient. 

Energy conservation yields that the total power output must be the same as the sum of the heat currents, $P=IV=-J_L-J_R$. Note that for the two-terminal setup considered here the power can be generated at either reservoir, so it is easier to label the heat currents as belonging to the hot or cold reservoir, respectively, $J_h$ and $J_c$. 
That is, for $T_L \gtrless T_R$ we have $T_h=T_{L,R}$, $T_c=T_{R,L}$ and, therefore, $J_h=J_{L,R}$ and $J_c=J_{R,L}$. 

With this notation we define three useful operational modes depending on the power generated or dissipated by the thermodynamic potentials~\cite{Rafa_2020}: 
A \textit{heat engine} (HE) produces electric power ($P>0$) from a temperature gradient $T_L\neq T_R$ such that heat is transferred from the hot to the cold reservoirs ($J_c<0$ and $J_h>0$); 
the \textit{heat pump} (HP) consumes power ($P<0$) to heat both reservoirs ($J_h, J_c < 0$); and 
the hybrid \textit{refrigerator-heat pump} (RP) is obtained when the device is powered ($P<0$) to heat the hot reservoir ($J_h<0$) and cool the cold one ($J_c>0$). In a two-terminal setup it is not possible to refrigerate one electrode without heating the other. 

\begin{figure}[t]
	\includegraphics[width=1.0\linewidth]{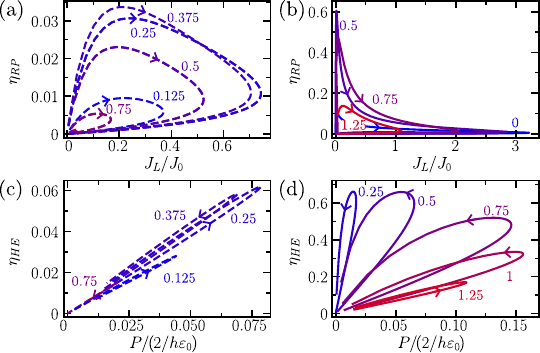}
	\caption{Lasso diagram of efficiency versus output power with implicit variable $eV$ traced clockwise for the refrigerator pump (a,b) and counterclockwise for the heat engine (c,d). Each line is computed for the indicated value of $U_C$. We set $m=0$ in (a,c), $m=m_0$ in (b,d), and $k_B T_R/m_0=1/8$, $k_B T_L/m_0=0.95/8$, $d/(\hbar v_F/m_0)=6$, $U_L=U_R=0$ and $\beta=0$ (out of plane magnetization) in all cases. }
	\label{fig:Lasso}
\end{figure}

Our aim is to explore the relevant operations for thermoelectric effects, HE and RP, since the heat pump represents the conventional behavior of an ohmic contact where an applied voltage ($P<0$) allows an electric current to flow between terminals, heating them ($J_{h,c}<0$) via Joule effect. 
A coefficient of performance (COP) for each operational mode is thus given by the ratio
\begin{equation}
	\text{COP}_\text{HE}=\frac{P}{J_h} , \quad
	\text{COP}_\text{HP}=\frac{J_h}{P} , \quad 
	\text{COP}_\text{RP}=\frac{J_c}{P} ,
\end{equation}
where, for example, $\text{COP}_\text{HE}$ represents the fraction of useful electric power $P$ generated by the heat flow $J_h$. Similarly, $\text{COP}_\text{RP}$ indicates how much cooling power $J_c$ can be obtained after supplying the system with power $P$. 
Introducing bounds to each COP, the thermodynamic efficiencies are 
\begin{equation}
	\eta_\text{HE} = \frac{1}{\eta_C} \frac{P}{J_h} , \quad
	\eta_\text{HP} = \eta_C \frac{J_h}{P}, \quad
	\eta_\text{RP} = \frac{1}{\eta_R} \frac{J_c}{P} ,
\end{equation}
where $\eta_C = 1 - T_c/T_h$ is the Carnot efficiency and $\eta_R = T_c/(T_h-T_c)$ the coefficient of performance for cooling reversibility. Note that $\eta_\text{RP}$ reaches maximum efficiency at $\Delta T \to 0$, while the $\eta_\text{HE}$ does it for $\Delta T \to \infty$. 

The efficiencies and their corresponding generated powers are shown in \cref{fig:Lasso} as \textit{lasso} diagrams, where each line corresponds to a different value of $U_C$ in the range between $U_C/m_0=0$ and $1.5$, and is computed by changing $eV$ clockwise (counter-clockwise) for $\eta_\text{RP}$ ($\eta_\text{HE}$) as indicated by the arrows. 

The cooling efficiency $\eta_\text{RP}$ at $m=0$, \cref{fig:Lasso}(a), reaches a maximum value of $\eta_\text{RP} \approx 0.03$ at around one third of the maximal cooling $J_0$. By turning on the magnetic region ($m\neq0$) the efficiency is increased tenfold [\cref{fig:Lasso}(b)], with maximum values of $20$-$60\%$ for up to half of the maximum cooling. 
Moreover, the cooling power can surpass $J_0$ ($J_L\sim 2$-$3 J_0$) at the cost of reducing efficiency to below $10\%$. 

The heat engine efficiency $\eta_\text{HE}$ for $m=0$ [\cref{fig:Lasso}(c)], reaches its maximum value $\eta_{HE}\approx 0.06$ at a maximum power output of up to $P/(2/h\epsilon_0)\approx0.08$, both at $U_C/m_0\sim0.25$. 
Again, the magnetic effect drastically enhances the efficiency, see \cref{fig:Lasso}(d). 
The heat engine efficiency is now large, $\eta_\text{HE}\approx0.5$, for a wide range of $U_C/m_0\approx0.25$-$0.75$, while the maximal power outputs are doubled compared to the non-magnetic case, $P/(2/h\epsilon_0)\approx0.01$-$0.15$. 

\begin{figure}[t]
	\includegraphics[width=0.99\linewidth]{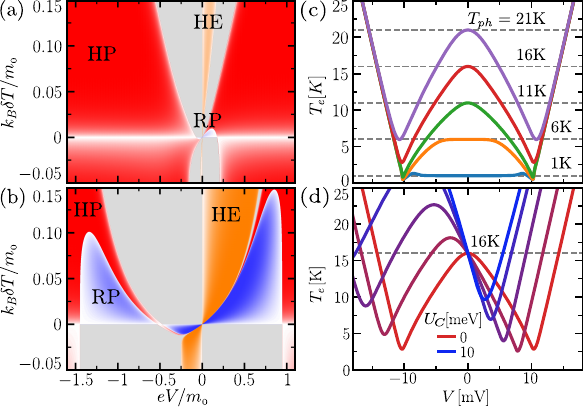}
	\caption{
		(a,b) Operational modes as a function of the thermodynamic potentials $eV$ and $\delta T$ for $m=0$ (a) and $m=m_0$ (b). Darker (brighter) colors indicate higher (lower) efficiency, and the configurations with no operational mode are shown in gray. We set $k_BT_L/m_0=0.6$, $\beta=\pi/4$, $U_C/m_0=1/4$ and $d=3 \hbar v_F/m_0$. 
		(c,d) Electron temperature $T_e$ at different phonon temperatures $T_\text{ph}$ for $d=500\text{~nm}$, $m=10\text{~meV}$, $\beta=0$, and $U_C=0$ (c) or $U_C=0$, $2.5$, $5$, $7.5$, and $10$~meV at $T_\text{ph}=16$~K (d). 
		}
	\label{fig:MapHBEQ}
\end{figure}

In \cref{fig:MapHBEQ}(a,b), we explore how the different operational modes appear as we take the system out of equilibrium either by an applied voltage $eV$ or a temperature gradient $\delta T$. Color grading indicates the efficiency of each operation, with darker colors for higher efficiencies. 
In the absence of magnetic barrier, \cref{fig:MapHBEQ}(a), the system mostly operates as a conventional heat pump, with very narrow HE (yellow) and RP (blue) regions. 
By contrast, for $m=m_0$ in \cref{fig:MapHBEQ}(b) the heat engine mode (yellow regions) easily appears for finite temperature gradients and requires a larger bias (\textit{thermovoltage}) to be suppressed, as compared to the $m=0$ case. The refrigerator pump operation (blue regions) is also greatly enhanced at low voltages $eV/m_0\lesssim1$. Moreover, the nonreciprocal transport in our setup favors cooling in one direction ($\delta T>0$). 

\subsection{Effective electron temperature} 

Thus far we only considered the electronic surface state, but in realistic setups at finite temperature the contribution from phonons can be important.  
We introduce heat dissipation, enabling an estimate of the maximum cooling temperature the device can obtain, by including interactions between electrons and phonons described by ensembles with temperatures $T_e$ and $T_\text{ph}$, respectively~\cite{Wellstood1994, Chen_2012, Bercioux2018, karimi_2020,Vischi2020May,Wiesner_2022}. 
The heat balance equation describes the exchange of heat between electron and phonons baths
\begin{equation}\label{eq:hbeq}
	\frac{\Sigma \mathcal{V}}{J_0} (T_e^\delta - T_\text{ph}^\delta) + \frac{J_L}{J_0}(T_e,T_\text{ph},eV) = 0, 
\end{equation}
where $\Sigma$ is a material dependent constant, $\mathcal{V}$ is the setup volume or area, and $\delta$ is a parameter usually considered to be $\delta=4$ in finite low-temperature two-dimensional materials~\cite{Sergeev2005Apr,Kivinen2003Sep,Giraud2012Jan}. 
We estimate the coupling as $\Sigma \mathcal{V}/J_0 \approx 1.41 \times 10^{5}$~meV~s$^{-1}$~K$^{-4}$~\cite{Chen_2012,Vischi2020May,Wiesner_2022}, with $m=10$ meV, using parameters corresponding to Bi$_2$Te$_3$~\cite{Wimmer2021Oct,Wang2023Aug} and a phonon bath area $\mathcal{V} \approx (10 ~\mu\text{m})^2$~\cite{Wiedenmann2017Oct,Charpentier2017Dec}. 
We then obtain an estimate for the electron temperature by solving \cref{eq:hbeq} numerically for $T_e$ at a given $T_\text{ph}$. The resulting electron temperature as a function of bias voltage $V$ is shown in \cref{fig:MapHBEQ}(c) for $U_C=0$ and in \cref{fig:MapHBEQ}(d) for different values of $U_C$. The phonon temperature used in each calculation is marked by dashed gray lines. 
We observe that the electron temperature sharply decreases for voltages below the gap edge ($eV/m\sim1$). Such an enhanced cooling power near the magnetic gap enables a reduction of $T_e$ from $21$K to $\sim 6$K, a reduction to $\sim30$\% of the phonon bath temperature. At lower temperatures electron cooling is still notable, with temperature reductions of up to $\sim10$\%. Biasing the central region ($U_C\neq 0$) slightly increases the cooling effect, which becomes asymmetric with the voltage due to the nonreciprocal transport effect, see \cref{fig:MapHBEQ}(d). 
We note that the cases with $\delta=4$ and $\delta=5$ as very similar, see \cref{sec:app-C}. 
The stark suppression of the electron temperature with respect to the phonon temperature for two-dimensional Dirac fermions is remarkable, although we note that the coupling constant $\Sigma$ we are using for surface states of \glspl{ti} is smaller than for other two-dimensional materials like graphene~\cite{Paolucci2024Feb,Chen_2012}. 

\section{Conclusions\label{sec:conc}}
We have proposed a mesoscopic thermoelectric device based on the surface states of a \gls{ti}, where a magnetic barrier defines a ballistic \gls{nfn} junction. Our theoretical analysis shows that this platform offers strong electrical tunability via gate control and supports efficient thermoelectric power generation and refrigeration. 
The unique spin-momentum locking of the surface states, combined with the magnetic barrier, gives rise to pronounced nonreciprocal transport, including diode-like charge and heat flow. 
Such an interplay between magnetization and spin–orbit coupling creates transport gaps in the \gls{nfn} junction, arising from either a spectral gap for out-of-plane magnetization ($m_z$) or momentum-space shifts for in-plane ones ($m_y$). 
In both cases, gate control over electrostatic potentials allows precise tuning of the conductance gap and transport asymmetry. 
%
We predicted nonreciprocal transport, including asymmetric cooling and electric diode effects, and demonstrated that the ballistic \gls{nfn} junction can operate as an efficient nanoscale refrigerator and heat engine while generating substantial output power. 
%
We have also evaluated the cooling performance by incorporating phonon contributions, confirming the relevance of our predictions for realistic quasi-two-dimensional materials, such as graphene or Bi-based topological insulators. 
These findings highlight the potential of topological surface states as a versatile platform for nanoscale thermal management.

\acknowledgments
We thank J.~Balduque and R.~S\'anchez for insightful discussions. 
P.M. and P.B. acknowledge support by the Spanish CM ``Talento Program'' project No.~2019-T1/IND-14088 and No.~2023-5A/IND-28927, the Agencia Estatal de Investigaci\'on project No.~PID2020-117992GA-I00 and No.~CNS2022-135950 and through the ``María de Maeztu'' Programme for Units of Excellence in R\&D (CEX2023-001316-M). 
B.L. acknowledges financial support from the National Natural Science
Foundation of China (project 12474049). 
B.S. acknowledges funding by the Deutsche Forschungsgemeinschaft
(DFG, German Research Foundation) Project No.~278162697-SFB 1242. 
Y.T. acknowledges financial support from JSPS with Grants-in-Aid for Scientific Research (KAKENHI Grants Nos. 23K17668, 24K00583, 24K00556, 24K00578, 25H00609 and 25H00613). 

\appendix

\section{Scattering amplitudes\label{sec:app-A}}

In this Appendix we show the details about matching the scattering states to compute the reflection and transmission amplitudes. For simplicity, we only consider the case of a particle injected from the left normal region. 

\begin{widetext}
Using \cref{eq:spinor-N} for the states in the normal regions and \cref{eq:spinor-C} for the magnetic one, the matching at the $x=0$ and $x=d$ interfaces takes the form
	\begin{equation}
		\psi_{L+}(0,\theta_L) + r_{LL} \psi_{L-}(0,\theta_L) =  a \psi_{C+}(0,\theta_C) + b \psi_{C-}(0,\theta_C) , \quad
		t_{RL} \psi_{R+}(d,\theta_R) =  a \psi_{C+}(d,\theta_C) + b \psi_{C-}(d,\theta_C) ,
	\end{equation}
with $r_{LL}$ and $t_{RL}$ respectively the reflection and transmission amplitudes. 

We solve this set of equations imposing that $U_L=U_R$, so that $\theta_L=\theta_R\equiv \theta$, as it is assumed for most of our results. The generalization for any values of $U_{L,R}$ is straightforward. 
The resulting amplitudes are
\begin{subequations}
	\begin{align}
		r_{LL} ={}& \frac{
		\sin (k_Cd) \me^{i\theta} (\sin\theta-\sin\theta_C)
		}{
		\cos (k_Cd) \cos\theta\cos\theta_C - i \sin (k_Cd) (1-\sin\theta\sin\theta_C)
		}, \\
		t_{RL} ={}& \frac{
		\me^{k_Cd} \cos\theta \cos\theta_C
		}{
		\cos (k_Cd) \cos\theta\cos\theta_C - i \sin (k_Cd) (1-\sin\theta\sin\theta_C)
		}.
	\end{align}
\end{subequations}

As a result, the reflection ($R$) and transmission ($T$) probabilities are
\begin{subequations}
	\begin{align}
			R= |r_{LL}|^2 ={}& \frac{
			\sin^2 (k_Cd) (\sin\theta-\sin\theta_C)^2
		}{
			\cos^2 (k_Cd) \cos^2\theta\cos^2\theta_C + \sin^2 (k_Cd) (1-\sin\theta\sin\theta_C)^2
		} , \\
			T= \mathrm{Re} \left(\frac{k_R}{k_L}\right) |t_{RL}|^2 ={}& \frac{
			\cos^2\theta \cos^2\theta_C
		}{
			\cos^2 (k_Cd) \cos^2\theta\cos^2\theta_C + \sin^2 (k_Cd) (1-\sin\theta\sin\theta_C)^2
		} .
	\end{align}	
\end{subequations}
\end{widetext}

\section{Analysis of transport gaps\label{sec:app-B}}

\begin{figure*}[ht!]
	\includegraphics[width=0.95\linewidth]{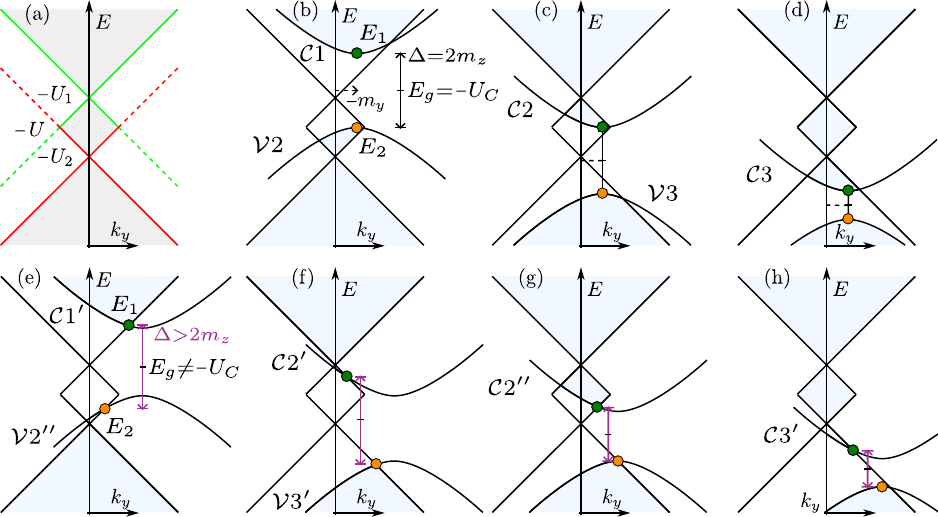}
	\caption{(a) Dispersion of $L$ and $R$ bands when $U_L\neq U_R$, with overlapping modes shown in gray. (b)-(h) Examples of conditions $\mathcal{C}1$ (b), $\mathcal{C}2$ (c), $\mathcal{C}3$ (d), $\mathcal{C}1'$ (e), $\mathcal{C}2'$ (f), $\mathcal{C}2''$ (g), and $\mathcal{C}3'$ (h), with available modes in blue. The lowest (highest) transmitting mode for the conduction (valence) band in region $C$ is marked by a green (orange) dot at energy $E_1$ ($E_2$). Vertical lines indicate the gap amplitude $\Delta$, centered at $E_g$, and horizontal dashed lines the central band displacement by $m_y$. In all cases, $k_x=0$. }
	\label{fig:bands-app}
\end{figure*}

We now analyze how the so-termed \textit{transport gap} appears in the conductance. As explained in the main text, the out-of-plane magnetization $m_z\neq0$ and $m_y=0$ ($\beta=0$), behaves like a mass term that gaps the spectrum of the central region around $U_C$. As a result, the conductance also features a gap. However, the emergence of a transport gap for the in-plane magnetization with $\beta=\pi/2$, or any situation with $0<\beta\leq\pi/2$, is not so straightforward. 
This phenomenon is explained by finding the intersection between available modes in the bands of regions $L$, $C$ and $R$. As we detail now, for some combinations of parameters a subset of modes can not be simultaneously available in $L$, $C$ and $R$, leading to the transport gap. 

In the general case, the relative position of the bands is determined by the parameters $U_{L,C,R}$ and $m_{y,z}$. 
Before even considering the central region, transport is only possible for a given energy when there are available states on both outer regions. 
Such a $LR$-\textit{intersection} is defined as the set of momenta that enables real solutions to the equations 
\begin{equation*}
	-U_L\pm \sqrt{k_x^2+k_y^2}=-U_R\pm \sqrt{k_x^2+k_y^2} ,
\end{equation*}
shown as a gray shaded region in \cref{fig:bands-app}(a) for the case with $k_x=0$. 
If the central region was completely transparent, the $LR$-intersection determines the maximum conductance. 
For a general description, we define $U_1= \min(U_L, U_R)$ and $U_2 = \max(U_L, U_R)$, so that $U=(U_L+U_R)/2=(U_1+U_2)/2$ and $\delta U=|U_R-U_L| = U_2 - U_1\geq0$. 
We then identify three important parts of the $LR$-intersection [\cref{fig:bands-app}(a)]: the \textit{upper} region above $-U_1$, the \textit{lower} one below $-U_2$; and the \textit{diamond} at $-U_2<E<-U_1$. 

We now study how the dispersion in the central region reduces the number of available states by analyzing the overlap between the bands in region $C$ and the $LR$-intersection. 
For simplicity, we assume $m_{y,z} \geq 0$. 
The central region has a gap of size $2m_z$ centered at $-U_C=(E_c+E_v)/2$, with $E_{c(v)} = -U_C +(-) m_z$ the bottom (top) of the conduction (valence) band in $C$. 
The blue shaded regions in \cref{fig:bands-app}(b) indicate the available modes for transport, that is, the points in $(E,k_y)$ space common to the three regions $L$, $C$, and $R$. After overlapping region $C$ with the $LR$-intersection we find an effective gap $\Delta= E_1- E_2$ centered at $E_g= (E_1+E_2)/2$, where $E_{1}$ and $E_{2}$ respectively indicate the lowest available mode in the conduction band (green dot) and the highest one in the valence band (orange dot). Note that for the case in \cref{fig:bands-app}(b) we have that $E_1=E_c$, $E_2=E_v$ and, thus, $\Delta=2m_z$ and $E_g=U_C$. However, as we detail below, this is not the general case. 

When we indeed have that $E_1=E_c$ and $E_2=E_v$ we have three configurations for both the conduction ($\mathcal{C}$) and valence ($\mathcal{V}$) bands on the central region. 
A sufficient but not necessary condition for this situation corresponds to $m_y=0$. 
Tracking the edge of each band, that is, the bottom of the conduction band and the top of the valence one, we find ($m_{y,z} \geq 0$): 
\begin{enumerate}
	\item The band edge is in the upper region when
	\begin{align}
		\label{eq:app-C1}
		\mathcal{C}1\text{: } \quad & -U_1 + m_y < -U_C + m_z , \\
		\label{eq:app-V1}
		\mathcal{V}1\text{: }\quad & -U_1 + m_y < -U_C - m_z .
	\end{align}
	\item The band edge is inside the diamond when 
	\begin{align}
		\label{eq:app-C2}
		\mathcal{C}2\text{:} \quad & -U_1 - m_y > -U_C + m_z > -U_2 + m_y , \\
		\label{eq:app-V2}
		\mathcal{V}2\text{:}\quad & -U_1 - m_y > -U_C - m_z > -U_2 + m_y . 
	\end{align}	
	\item The band edge is in the lower region when
	\begin{align}
		\label{eq:app-C3}
		\mathcal{C}3\text{: } \quad & -U_2 - m_y > -U_C + m_z, \\
		\label{eq:app-V3}
		\mathcal{V}3\text{: }\quad & -U_2 - m_y > -U_C - m_z . 
	\end{align}
\end{enumerate}

An example of condition $\mathcal{C}1$ is showcased in \cref{fig:bands-app}(b), see green dot. In the same panel, the top of the valence band (orange dot) is inside the diamond, which corresponds to condition $\mathcal{V}2$. Similarly, the bottom of the conduction band is inside the diamond in \cref{fig:bands-app}(c) (condition $\mathcal{C}2$) and in the lower region in \cref{fig:bands-app}(d) (condition $\mathcal{C}3$). These two panels show examples of $\mathcal{V}3$ for the orange dot. Note that in all these cases the lowest transmitting mode above the gap has energy $E_1=E_c$, while below the gap we have $E_2=E_v$ (see blue regions). 

More generally, when $E_1\neq E_c$ or $E_2\neq E_v$ we can extend the previous conditions ($m^2=m_y^2+m_z^2$): 
\begin{enumerate}
	\item[1'.] The band edge is in the upper region when
	\begin{align}
		\label{eq:app-C1p}
		\mathcal{C}1'\text{: } \quad & m_z - m_y < U_C - U_1 < m , \\
		\label{eq:app-V1p}
		\mathcal{V}1'\text{: }\quad & -m_z - m_y < U_C - U_1 < - m .
	\end{align}
	\item[2'.] The band edge is inside the diamond, cutting the top band, if 
	\begin{align}
		\label{eq:app-C2p}
		\mathcal{C}2'\text{: } \, & |U| + \sqrt{(|U| + m_y)^2 + m_z^2} > U_C - U_1 >  m , \\
		\label{eq:app-V2p}
		\mathcal{V}2'\text{: }\, & |U| - \sqrt{(|U| + m_y)^2 + m_z^2} > U_C - U_1 > - m . 
	\end{align}	
	\item[2''.] The band edge is inside the diamond, intersecting the bottom band, when 
	\begin{align}
		\label{eq:app-C2pp}
		\mathcal{C}2''\text{: } \, & \sqrt{(U + m_y)^2 + m_z^2} - |U| < U_C - U_2 < m, \\
		\label{eq:app-V2pp}
		\mathcal{V}2''\text{: }\, & -\sqrt{(U+ m_y)^2 + m_z^2} - |U| < U_C - U_2 < - m . 
	\end{align}	
	\item[3'.] The band edge is in the lower region when
	\begin{align}
		\label{eq:app-C3p}
		\mathcal{C}3'\text{: } \quad & m_y + m_z < U_C - U_2 < m , \\
		\label{eq:app-V3p}
		\mathcal{V}3'\text{: }\quad & m_y - m_z < U_C - U_2 < - m . 
	\end{align}
\end{enumerate}

A finite $m_y$ displaces the bands in the $C$ region and, as a result, the edges of the band gap, $E_{c,v}$, may not correspond to transmitting modes anymore. For example, the green dot in \cref{fig:bands-app}(e) represents a situation where the bottom of the conduction band is outside the $LR$-intersection. The lowest-energy transmitting state in the conduction band is now at $E_1>E_c$ inside the upper region (condition $\mathcal{C}1'$). At the same time, the orange dot is at $E_2<E_v$ in the lower region, representing condition $\mathcal{V}3'$. 
\Cref{fig:bands-app}(f) and \ref{fig:bands-app}(g) showcase examples of conditions $\mathcal{C}2'$ and $\mathcal{C}2''$, respectively (green dots), and \cref{fig:bands-app}(h) is an example of condition $\mathcal{C}3'$. 

Another consequence of the displaced central bands is that the effective gap no longer coincides with the gap of region $C$, $\Delta\neq 2m_z$. Indeed, for all examples in the bottom panels of \cref{fig:bands-app} the effective gap is bigger than the band gap and no longer centered around $U_C$. 
This explains how we can have a transport gap $\Delta\neq0$ for a finite $m_y$ even if $m_z=0$. 

The analysis of the emergent transport gap allows us to determine the asymmetry in the conductance that leads to the strong thermoelectric effects discussed in the main text. 
We now enumerate some examples of an asymmetric conductance: 
\begin{enumerate}
	\item Any combination of $\mathcal{C}1-\mathcal{C}3$ with $\mathcal{V}1-\mathcal{V}3$ results in a transport gap equivalent to the band gap in region $C$ ($\Delta= 2m_z$ located at $E_g=-U_C$). 
	\item The combinations between $\mathcal{C}1-\mathcal{C}3$ and primed conditions $\mathcal{V}1'$, $\mathcal{V}2'$, $\mathcal{V}2''$ and $\mathcal{V}3'$ yield a gap $\Delta>2m_z$ and no longer centered around $-U_C$. 
	\item For the specific combinations ($\mathcal{C}1',\mathcal{V}2'$) or ($\mathcal{C}2'',\mathcal{V}3'$), if $m_z=0$ the transport gap is centered at $E_g = |U_1-U_C|/2$ or $E_g = |U_2-U_C|/2$, respectively. If $m_z\neq 0$, and if $U_C$ coincides with either $U_1$ or $U_2$, then $E_g=U_C$ and the gap will always be $\Delta=m^2/m_y \geq 2m_z$. 
	\item An interesting situation corresponds to the \textit{bipolar configuration} with $U_1-U_C= U_C-U_2$, or simply $U_C=U$. We are reduced to the combinations ($\mathcal{C}1,\mathcal{V}3$) or ($\mathcal{C}2,\mathcal{V}2$) with $\Delta=2m_z$, and ($\mathcal{C}1',\mathcal{V}3'$) or ($\mathcal{C}2',\mathcal{V}2''$) where $\Delta>2m_z$. The latter cases feature a gap of magnitude 
	\begin{equation}
		\Delta_\text{bp} = \frac{4 m^2 + 4m_y \delta U + \delta U^2}{4m_y + 2 \delta U},
	\end{equation}
	 with $\delta U = U_1-U_2 > 0$ and $\delta U \geq 2m_y $. 
	 Note that the gap size is dependent on the degree of bipolarity on the system ($\delta U$). When $\delta U=0$ the gap is $\Delta=m^2/m_y$. On the other hand, for $\delta U \neq 0$ and in-plane magnetization ($\beta=\pi/2$) we have
	\begin{equation}
		\lim_{m_z\to 0} \Delta_\text{bp} = m_y + \delta U/2.
	\end{equation}
	Note that the gap is maximal at $2m_y$ when $\delta U =2m_y$ corresponding to $U_1=m_y$ and $U_2=-m_y$. 
\end{enumerate}

\begin{figure}[t]
	\includegraphics[width=0.95\linewidth]{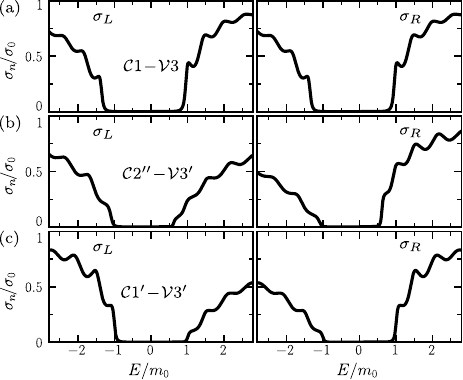}
	\caption{Conductance $\sigma_L$ (left column) and $\sigma_R$ (right column) as a function of energy. 
	(a) Conductance under conditions $\mathcal{C}1$ and $\mathcal{V}3$ with $U_{L,R}=0$, $U_C/m_0=0.2$ and $\beta=0$. 
	(b) Conductance under conditions $\mathcal{C}2''$ and $\mathcal{V}3'$ with $U_L=-U_R=U_C=0.4m_0$ and $\beta=\pi/2$. 
	(c) Conductance under conditions $\mathcal{C}1'$ and $\mathcal{V}3'$ with $U_L=-U_R=0.6m_0$, $U_C=0$ and $\beta=\pi/4$. 
	In all cases $d=4 \hbar v_F/m_0$. 
	}
	\label{fig-app:condUR}
\end{figure}

The transport asymmetry in example 1 above appears for $U_C\neq 0$, but the transport gap is mostly given by the band gap in the central region. Consequently, the conductance on the left and on the right regions, $\sigma_{L,R}$, coincides as long as the dopings $U_{L,R}$ are the same or symmetrical with respect to $U_C$. This situation, depicted in \cref{fig-app:condUR}(a), leads to thermoelectric effects but not to nonreciprocal currents, since $\sigma_n(E)\neq \sigma_n(-E)$, for $n=L,R$, but $\sigma_L(E)=\sigma_R(E)$. 

To reach a strong nonreciprocal transport, it is helpful to enhance the asymmetry in the transport gap (examples 2 and 3) to fulfill $\sigma_n(E)\neq \sigma_n(-E)$ together with $\sigma_L(E) \neq \sigma_R(E)$. An example of this situation is given by the combination of $\mathcal{C}2''$ and $\mathcal{V}3'$, see \cref{fig-app:condUR}(b). 
Now, $\sigma_L$ and $\sigma_R$ feature a similar gap, but the transmitting energies are very different. 

Finally, the bipolar configuration (example 4), represented by the conditions $\mathcal{C}1'$ and $\mathcal{V}3'$ in \cref{fig-app:condUR}(c), results in the connection $\sigma_L(E) = \sigma_R(-E)$, while keeping $\sigma_n(E)\neq \sigma_n(-E)$. This situation is useful for both thermoelectric effects and nonreciprocal transport. 

\section{Calculation of effective electron temperature\label{sec:app-C}}

\begin{figure*}[t]
	\includegraphics[width=0.95\linewidth]{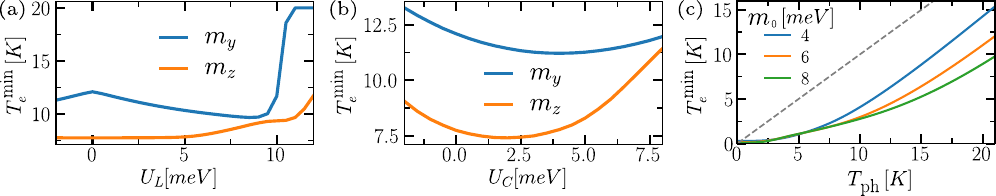}
	\caption{
		Minimum electron temperature $T_e^\text{min}$ for a positive voltage as a function of (a) $U_L=-U_R$ at $U_C=0$, (b) $U_C$ at $U_{L,R}=0$, and (c) $T_\text{ph}$ for different values of $m$. We set $m=10$ meV and $T_\text{ph}=20$ K for (a,b), and $U_{L,C,R}=0$ for (c). The dashed line in (c) indicates the original electron temperature before cooling (i.e., $T_\text{ph}$). 
	}
	\label{fig-app:maps}
\end{figure*}

Here, we further justify \cref{eq:hbeq} in the main text, that is, the heat balance equation between surface state electrons and the phonon bath. 
We follow Ref.~\cite{Wiesner_2022} where heat transport was measured for a \gls{ti}. 
This work identified a $T^4$ contribution from electron-phonon coupling and a $T^2$ one from electron-electron interactions. Both contributions can be taken into account by 
\begin{equation}\label{eq:hbeq2}
	c_2 (T_e^2 - T_\text{ph}^2) + c_4 (T_e^4 - T_\text{ph}^4) + J_L(T_e,T_\text{ph},eV) = 0, 
\end{equation}
with $c_2 = 6.24 \times 10^7 \text{~meV s}^{-1} \text{~K}^{-2}$ and $c_4 = 6.87 \times 10^5 \text{~meV s}^{-1} \text{~K}^{-4}$ and $J_L$ being the cooling of the electronic reservoir~\cite{Wiesner_2022}. 
Solving \cref{eq:hbeq2} and, analogously, \cref{eq:hbeq} of the main text, involves finding the temperature $T_e$ that satisfies the equation for a given cooling power $J_L$. This is achieved at a reference temperature $T_\text{ph}$. 
After numerically solving \cref{eq:hbeq2} we find that for $T_\text{ph}<30$ K the $c_2$ term has a negligible effect on the electron temperature. We thus ignored this term in \cref{eq:hbeq} in the main text, although we include it here for clarity. The values of $\Sigma V$ in the main text, estimated from bibliography~\cite{Chen_2012,Vischi2020May,Wiesner_2022,Wimmer2021Oct,Wang2023Aug,Wiedenmann2017Oct,Charpentier2017Dec}, are similar in magnitude to the $c_4$ coefficient yielding very similar electron temperatures.  

As we showed in the main text, cf. \cref{fig:MapHBEQ}(c,d), the solution to \cref{eq:hbeq2} for a given $T_\text{ph}$ is symmetric with the voltage when $U_C=0$ and asymmetric for $U_C\neq0$, respectively featuring two minima or a single minimum. For positive voltages, we label this minimum electron temperature $T_e^\text{min}$. In the main text we only analyzed the case with $U_L=U_R=0$. We now explore the bipolar situation with $U_L=-U_R$ and $U_C=0$ in \cref{fig-app:maps}(a) for in-plane ($\beta=\pi/2$) and out-of-plane ($\beta=0$) magnetizations. The minimum electron temperature is stable for a large range of bipolar dopings, with the cooling for the out-of-plane case always slightly better since the gap edge is better defined. 

The asymmetric cooling for $U_C\neq0$ shown in \cref{fig:MapHBEQ}(d) indicates there is a maximum electron cooling for a given $U_C>0$. We explore this maximum cooling (minimum of $T_e^\text{min}$) in \cref{fig-app:maps}(b). The cooling effect is indeed maximized for $U_C\sim2.5$~meV ($U_C\sim5$~meV) when only $m_z$ ($m_y$) is finite. 

Finally, in \cref{fig-app:maps}(c) we explore the electron cooling effect at different strengths of the magnetic barrier for similar values as in \cref{fig:MapHBEQ}(c). The original electron temperature before cooling, $T_e=T_\text{ph}$, is shown as a dashed line. At low temperatures $T_\text{ph} \lesssim 7.5$~K, the cooling effects are similar for the magnetizations considered. At higher temperatures, but still below the regime where the quadratic term in \cref{eq:hbeq2} becomes relevant, the evolution of $T_e^\text{min}$ with the temperature of the phonon bath becomes almost linear, and the electron cooling is increased with the strength of the magnetic barrier $m_0$. 


%

\end{document}